# Intelligent Road Inspection with Advanced Machine Learning; Hybrid Prediction Models for Smart Mobility and Transportation Maintenance Systems

**Nader Karballaeezadeh** [1], **Farah Zaremotekhases** [2], **Shahaboddin Shamshirband** [3,4,*], **Amir Mosavi** [5,6,7,*], **Narjes Nabipour** [8,*], **Peter Csiba** [9], **Annamária R. Várkonyi-Kóczy** [9,10]

[1] Department of Civil Engineering, Shahrood University of Technology, Shahrood 3619995161; n.karballaeezadeh@shahroodut.ac.ir

[2] Department of Construction Management, Louisiana State University, Baton Rouge, LA 70803, USA; Fzarem1@lsu.edu

[3] Department for Management of Science and Technology Development, Ton Duc Thang University, Ho Chi Minh City, Vietnam

[4] Faculty of Information Technology, Ton Duc Thang University, Ho Chi Minh City, Vietnam; shahaboddin.shamshirband@tdtu.edu.vn

[5] School of the Built Environment, Oxford Brookes University, Oxford OX3 0BP, UK; a.mosavi@brookes.ac.uk

[6] Institute of Structural Mechanics, Bauhaus University Weimar, D-99423 Weimar, Germany; amir.mosavi@uni-weimar.de

[7] Faculty of Health, Queensland University of Technology, 130 Victoria Park Road, Brisbane City, QLD 4059, Australia; a.mosavi@@qut.edu.au

[8] Institute of Research and Development, Duy Tan University, Da Nang 550000, Vietnam; narjesnabipour@duytan.edu.vn

[9] Department of Mathematics and Informatics, J. Selye University, 94501 Komarno, Slovakia; csibap@ujs.sk

[10] Kalman Kando Faculty of Electrical Engineering, Obuda University, 1034-Budapest, Hungary; varkonyi-koczy@uni-obuda.hu, amir.mosavi@kvk.uni-obuda.hu (A.M)

**Abstract:** Prediction models in mobility and transportation maintenance systems have been dramatically improved through using machine learning methods. This paper proposes novel machine learning models for an intelligent road inspection. The traditional road inspection systems based on the pavement condition index (PCI) are often associated with the critical safety, energy and cost issues. Alternatively, the proposed models utilize surface deflection data from falling weight deflectometer (FWD) tests to predict the PCI. Machine learning methods are the single multi-layer perceptron (MLP) and radial basis function (RBF) neural networks as well their hybrids, i.e., Levenberg-Marquardt (MLP-LM), scaled conjugate gradient (MLP-SCG), imperialist competitive (RBF-ICA), and genetic algorithms (RBF-GA). Furthermore, the committee machine intelligent systems (CMIS) method was adopted to combine the results and improve the accuracy of the modeling. The results of the analysis have been verified through using four criteria of average percent relative error (APRE), average absolute percent relative error (AAPRE), root mean square error (RMSE), and standard error (SD). The CMIS model outperforms other models with the promising results of APRE=2.3303, AAPRE=11.6768, RMSE=12.0056, and SD=0.0210.

## 1. Introduction

In road transportation, pavement plays a vital role as the part of the road that is in direct contact with vehicles. Users' judgment about the quality of road service is primarily predicated upon pavement conditions. The Maintenance, Rehabilitation, and Reconstruction (MR&R) program of pavement network is a multidimensional decision-making process that takes into account several considerations. Highway agencies generally focus on two issues: maximizing the efficiency of the pavement network or minimizing agency costs [1]. Both of these issues require the estimation of operating conditions of the pavement network to set up pavement management and maintenance plans.

Detection of pavement surface distresses is essential before setting up a maintenance plan to determine the pavement operating conditions, as pavement network maintenance operations are arranged based on the state of such distresses [2]. Table 1 shows an instance of pavement network maintenance operations based on pavement status.

**Table 1.** Maintenance program in roads [2]

| Damage (%) | Condition | Maintenance Program |
|---|---|---|
| < 6 | Good | Routine Maintenance |
| 6-11 | Moderate | Minor Rehabilitation |
| 11-15 | Light Damage | Major Rehabilitation |
| > 15 | Heavy Damage | Reconstruction |

Pavement maintenance has attracted growing attention of pavement engineers in recent years. Evaluation of pavement conditions is the most important factor for the effective and economical maintenance of the pavement network that can lead to the promotion of service life [3]. The condition of an in-service pavement is assessable in two categories including functional and structural. Both functional and structural conditions play an important role in pavement management at the network-level[4]. In most Pavement Management Systems (PMSs), non-structural indices such as Pavement Condition Index (PCI) are used as pavement indicators to select treatments [3,5] while ignoring the structural conditions of pavement [6]. It has recently been proven that there is a statistical relationship between functional and structural conditions [7]. Hence in recent years, various agencies around the world have attempted to use indices of structural capacity in PMS and decision-making processes [6].

A very common index in the PMSs is the PCI, which was developed by the US Army Corps of Engineers in 1982. PCI is an indicator of surface functional condition and structural integrity [8]. After visual inspection of the pavement network, pavement engineers calculate PCI based on distress type, severity, and quantity. This index varies from zero for a virtually unusable pavement to 100 for a perfect pavement [9]. On the other hand, the assessment of structural conditions generally performed by non-destructive tests such as Falling Weight Deflectometer (FWD) [10-13]. In the FWD test, an impulsive load applies to the pavement surface for 25-30 ms and the surface deflections are recorded by seven (or more) sensors. The sensors measurements are analyzed by back-calculation software such as ELMOD and MODULUS and useful information, including overlay thickness, layers modulus, and remaining life is determined [14-16].

Nowadays, PCI calculation in many organizations is done using automated distress identification from digital images but some organizations are still using traditional methods for PCI determination. The traditional process of calculating PCI in a pavement segment involves the visual inspection of pavement. This type of inspection has always raised safety concerns among engineers. When an inspector is recording surface pavement distresses, the possibility of clashing into road traffic is relatively high. Another drawback of the PCI calculation process is concerned with potential human error in identifying and recording pavement distress. Human error can affect the accuracy of the calculated PCI. In this paper, the authors propose a novel method for estimating PCI in flexible pavements. In the proposed method, the PCI of a pavement segment can be calculated based on the surface deflections recorded in the FWD testing. Another major incentive of the authors was the inadequacy of studies on the link between PCI and pavement surface deflection in the FWD testing.

To implement the proposed method, the authors selected 236 pavement segments from Tehran-Qom freeway in Iran. First, PCI was calculated by inspecting all segments and recording surface distresses data. Then, all the segments were subjected to FWD testing and the average deflection of each segment was determined. After completing the database, the analysis stage was carried out with the help of Machine Learning (ML) techniques. ML techniques are a sub-category of computational intelligence. These techniques chiefly applied for function approximation, classification, pattern recognition, etc. [17]. ML include different methods such as ANN, RBF, SVM, etc. and many papers have been published on the application of these techniques in the PMS studies [18-29]. In this paper, Multi-Layer Perceptron (MLP) and Radial Basis Function (RBF) neural networks were used for data analysis. The optimization of MLP neural network was conducted by Levenberg-Marquardt (LM) and Scaled Conjugate Gradient (SCG) algorithms and RBF neural network was optimized by Genetic Algorithm (GA) and Imperialist Competitive Algorithm (ICA) algorithms. Therefore, the analytical methods used in this paper are MLP-LM, MLP-SCG, RBF-GA, and RBF-ICA. Finally, to obtain more accurate results, all four methods were merged with Committee Machine Intelligent Systems (CMIS) in a single model. The results of these five methods were assessed using Average Percent Relative Error (APRE), Average Absolute Percent Relative Error (AAPRE), Root Mean Square Error (RMSE) and Standard Error (SD) criteria to determine the accuracy of each method.

The proposed method enhances the safety of the PCI determination process by eliminating the field inspection of the pavement. Also, by removing the human factor from the inspection process, the potential human error is eradicated and the results is boosted. Furthermore, the use of FWD test results to estimate PCI provides to the overlapping and cost-effectiveness of the pavement network maintenance activities.

This paper is organized as follows. The second section reviews relevant studies in the literature. In the next section, the research methodology is presented. This section also includes an introduction to PCI and its calculation method, FWD test, the freeway understudy, and data analysis methods. At the end of this section, the validation criteria of the results are presented. The fourth section discusses the results and conclusions are drawn in the fifth Section.

## 2. Literature Review

In addition to the conventional method of determining PCI, several other attempts have been made to predict this index. PCI prediction methods can be broadly divided into three categories:
- PCI prediction methods based on other pavement quality indices
- PCI prediction methods based on pavement age
- PCI prediction methods based on pavement surface deflection

The first category represents the most frequent method used by other researchers. In this category, PCI of flexible pavement is determined based on other indices that manifest pavement quality. Surface pavement distresses have a direct relationship with other indicators of pavement distress, including roughness and driving quality. Table 2 depicts a number of methods in this category.

Table 2. PCI prediction methods based on other pavement quality indices

| Model | Equation | Description |
|---|---|---|
| Park et al. [30] | $\log(PCI) = 2 - 0.436 \log(IRI)$ | IRI: International roughness index. |
| Dewan and Smith [31] | $PCI = 153 - \dfrac{IRI}{0.0171}$ | IRI: International roughness index. |
| Arhin et al. [32] | $PCI = (A \times IRI) + K + \varepsilon$ | IRI: International roughness index, A and K: Regression coefficients, ε: Error |
| Korea institute of construction technology [33] | $NHPCI = ((0.004 x_{RD}) + (0.003 x_{CR}) + (0.0183 x_{IRI}) + 0.33)^{-2}$ | NHPCI: National highway pavement condition index, $x_{RD}$: Ruth depth (mm), $x_{CR}$: Crack ration (%), $x_{IRI}$: International roughness index (m/km). |
| Korea expressway corporation research institute [33] | $HPCI = 5 - 0.75 RD^{1.2} - 0.54 IRI^{0.8} - 0.9 \log(1 + SD)$ | HPCI: Highway pavement condition index, RD: Ruth depth (mm), IRI: International roughness index (m/km), SD: Surface distress (crack quantity converted to area) (m²). |
| Ningyuan et al. [34] | $PCI = DMI \times C_i \times \sqrt{0.1 \times RCI} \times 10$ | DMI: Distress manifestation index, $C_i$: Calibration coefficient for pavement type, RCI: Riding comfort index. |
| Khattak et al. [35] | $PCI = \text{Max} \begin{cases} 1.\,\text{Avg}\,(RNDM, ALCR, PTCH, RUFF, RUT) - 0.85 STD(RNDM, ALCR, PTCH, RUFF, RUT) \\ 2.\,\text{Min}\,(RNDM, ALCR, PTCH, RUFF, RUT) \end{cases}$ | RNDM: Random cracking index, ALCR: Alligator cracking index, PTCH: Patch index, RUFF: Roughness index, RUT: Rutting index. |

In the second category of PCI prediction methods, researchers focus on pavement age as a major prediction factor. The pavement age is directly linked to pavement distresses so that different types of distress are more likely to appear in an old pavement segment. Table 3 reveals examples of methods in this category.

Table 3. PCI prediction methods based on pavement age

| Model | Equation | Description |
|---|---|---|
| South Dakota Department of Transportation [36] | $PCI = a + (b \times age^c)$ | a: Maximum value of PCI, Age: Age of pavement (year), b: Slop coefficient of performance curve, c: Power coefficient for performance curve. |

| Oklahoma airfield pavement management system [37] | $PCI = a_0 + a_1x + a_2x^2 + \cdots + a_nx^n$ | $a_i$: Polynomial parameters, x: Pavement age, n: Polynomial order. |
|---|---|---|
| Michles [38] | $PCI = 71.09 + 27.42(\text{Treatment type}) - 4.07(\text{Age})$ | Treatment type: 0 for microsurfacing and 1 for thin overlay, Age: Pavement age (year). |

The third category of PCI prediction methods involves the pavement surface deflections in the FWD test. FWD is a device used to evaluate the structural capacity of pavements. The appearance of different types of surface distress on pavements and their expansion reflects the deterioration of the structural capacity of pavements. Given the above points, there is a mutual relationship between the FWD testing process and surface distresses of pavement. The paucity of research in this area was one of the reasons prompting the authors to investigate the relationship between pavement deflection data and the PCI index. One of the few studies that fall into this category of PCI prediction methods is the research undertaken by O'Brien et al. They developed a model for predicting PCI, which in addition to deflections of surface pavement, drew on traffic data, pavement age, and type of pavement[39]. Eq.1 shows the model proposed by O'Brien et al.

$$PCI = 96.6 - \Big[(0.000572 \times AGE^2 \times LPMTOT \times DIFF \times AREA) \\ + \Big(0.3062 \times AGE^{\frac{1}{4}} \times AGESOL^2 \times DIFF^2\Big) \\ + (0.00156 \times AGE^{\frac{1}{2}} \times AGETOT \times LPMTOT \times DIFF \times AREA)\Big] \quad (1)$$

where, AGE: Age of pavement since last overlay (yaer), LPMTOT: Log of weighted traffic total (veh/day), and DIFF: Normalized deflection basin slope,

$$DIFF = \frac{D_0 - D_{12}}{D_0} \quad (2)$$

where, $D_i$: Pavement surface deflection at i inches from center of loading plate in FWD test, and AREA: Area of FWD deflection basin at the high load level (in.$^2$/10$^3$),

$$AREA = 12(D_0 + D_{12}) \quad (3)$$

where, AGESOL: Age of pavement to last overlay (year), and AGETOT: Total age of pavement (year).

The data used by O'Brien et al. were obtained from Virginia in the United States. In this study, the pavement surface deflections data were collected using Dynatest 8000 FWD. Statistically, Eq.1 is moderately accurate, for a correlation coefficient ($R^2$) and a standard error ($\sigma$) of 0.586 and 6.88 were obtained, respectively[39].

Safety has always been a key factor in transportation engineering. As such, one major strength of the method proposed in this paper is that it eliminates the need for field inspection of pavement surface distresses, which significantly promotes inspection safety. On the other hand, whenever the human factor is involved in scientific processes, the possibility of error induced by inaccuracy and distraction cannot be ruled out. Thus, the accuracy of the PCI estimation process could be improved by eliminating the human factor. Another strength of this study lies in its application of FWD. Due to its very accurate simulation of traffic load, FWD is a valid test endorsed by all transportation agencies and is extensively used in many parts of the world for structural evaluation of pavements. Therefore, the simultaneous use of FWD testing for structural evaluation of the pavement network and PCI estimation contributes to the overlapping of maintenance activities and diminishes the consumption budget. The last achievement of this is concerned with its role in filling the research gap in this area, which could lay the ground for future research in this field.

## 3. Methodology

### 3.1. Pavement Condition Index (PCI)

One of the most common indices used to evaluate flexible pavement is PCI. Introduced by the US Army Corps of Engineers, this index is based on visuals inspection of pavement [40,41]. The PCI value is a number from 100 to zero, with 100 representing the best pavement conditions and zero indicating the worst pavement conditions. To calculate PCI in a pavement segment, initially, number 100 is assigned to that segment. Then, based on the type, extent, and severity of the pavement distresses, a Deduct Value (DV) is subtracted from until PCI is finally obtained[42]. Table 4 shows the relationship between the pavement status and the value of PCI.

**Table 4.** Rating scale of PCI [43]

| PCI | 0 - 10 | 10 - 25 | 25 – 40 | 40 – 55 | 55 – 70 | 70 – 85 | 85 - 100 |
|---|---|---|---|---|---|---|---|
| **Condition** | Failed | Serious | Very poor | Poor | Fair | Satisfactory | Good |

The process of PCI calculation in flexible pavements is summarized as follows [43,44]:
1. Determine the type, extent, and severity of pavement distresses.
2. Determining DV for each distress based on its corresponding curve. Figure 1 shows an example of such curves.

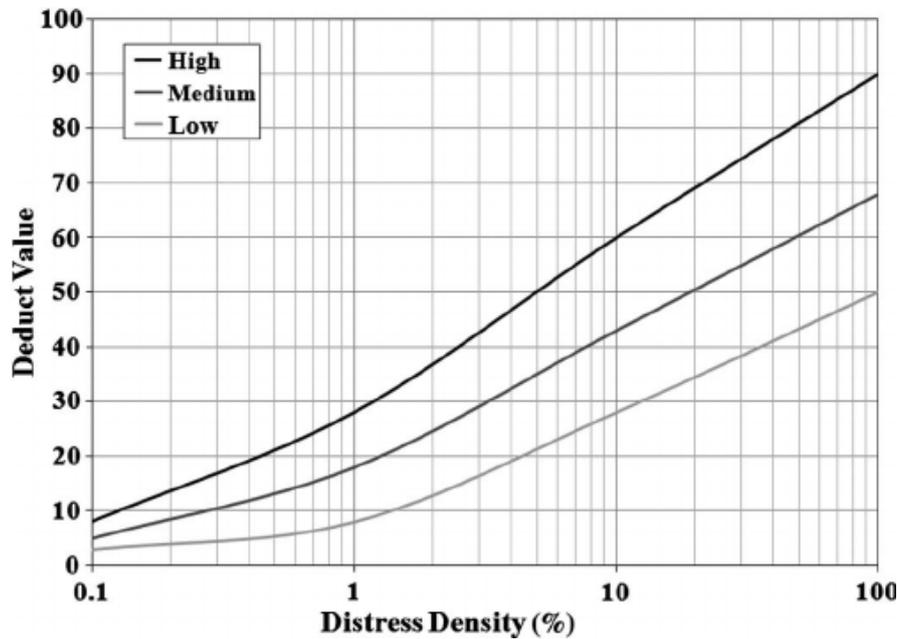

**Figure 1.** Typical deduct value curve [44]

3. Reducing the number of DVs to the maximum number allowed by Eq.4:

$$m_i = 1 + \frac{(100 - \text{HDV}) \times 9}{98} \tag{4}$$

where, $m_i$: Maximum allowable number of deduct values, and HDV: Greatest individual deduct value.

4. Determining the number of DVs greater than 2 (q).
5. Determining Total Deduct Value (TDV), which is the sum of DVs.
6. Determining Corrected Deduct Value (CDV) based on correction curves using q and TDV. Figure 2 shows an example of correction curves.

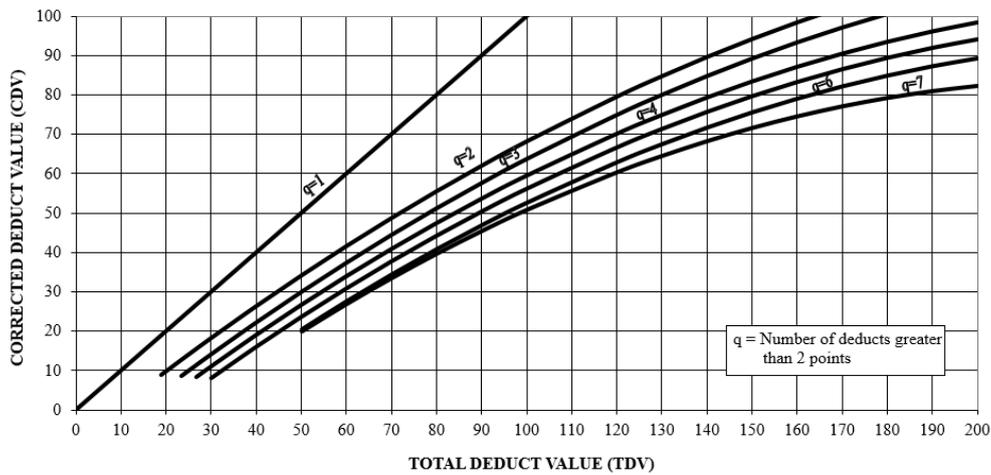

**Figure 2.** Typical corrected deduct value curve [44]

7. Decreasing the smallest DVs larger than 2 to 2.
8. Repeating steps 4 to 7 until q reaches 1.
9. Determining the maximum CDV and calculating the PCI using Eq.5:

$$\text{PCI} = 100 - \text{CDV}_{\max} \tag{5}$$

*3.2. Falling Weight Deflectometer (FWD)*

Structural evaluation of the pavement network is one of the requirements of pavement management systems and FWD is the most common test for structural evaluation of pavement[45]. This test is widely used by pavement engineers due to the desirable simulation of traffic load. In this experiment, a loading plate with a radius of 30 cm and 7 to 9 sensors installed at different distances from the center of the loading plate is placed on the pavement surface. The FWD applies a type of impulsive load to the pavement surface. To do so, a weight is dropped from a certain height on the loading plate[16]. Figure 3 shows the FWD load application system.

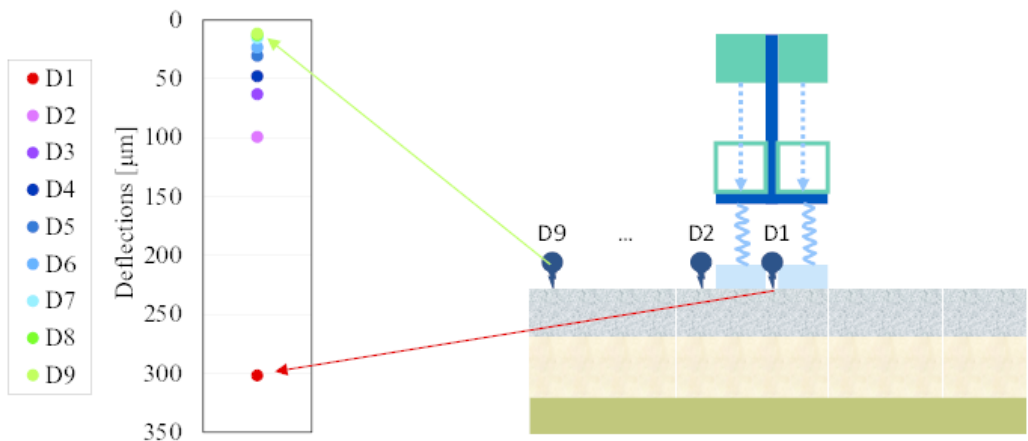

**Figure 3.** Loading mechanism in FWD [46]

After the load is applied to the pavement surface, it generates vertical deflections, which are recorded by the sensors. In this study, the resulting deflections are recorded by one geophone below the loading plate and six other geophones that are 20, 40, 60, 90, 120 and 150 cm away from the center of the loading plate. Deflections are transmitted to the central computer for later applications. Useful information such as remaining service life of the pavement, overlay thickness, and layers module can be obtained from the pavement surface deflections [16].

*3.3. Case study*

In this paper, 236 pavement segments were adopted from the Tehran-Qom freeway in Iran to implement the proposed theory. The understudy route is part of the artery between the capital and southern Iran, which is located in two provinces of Tehran and Qom. The freeway consists of 3 lanes in each direction with a width of 3.65 m for any lane. This freeway has a flexible pavement. A total of 236 pavement segments were selected from this freeway and the PCI was calculated as described in the subsection pavement condition index. After calculating PCI, a load was applied to the pavement using an FWD equipped with 7 pavement deflection recording sensors. These sensors recorded the mean deflection in all pavement segments.

*3.4. Analysis methods*

Artificial neural networks constitute a set of computational intelligence inspired by biological neural systems such as the human brain. Neural networks can be used to explore complex relationships between inputs and outputs of a system. Each neural network comprises two main elements: the processor elements (neurons and nodes) that process information, and weights, which are responsible for establishing connections between neurons [47]. The most common artificial neural networks are MLP and RBF, which have been used in this paper.

3.4.1. Multi-layer Perceptron (MLP) Neural Network

There are three types of layers in MLP. The first layer is the input layer, which is concerned with the input data. The second type of layer is the output layer that deals with the model output. Between the input and output layers, there are intermediate layers known as hidden layers. The number of neurons in the input layer is equal to the number of input variables, while the output is generally the parameter considered for the analysis. The number of hidden layers and neurons in each hidden layer is determined experimentally. Generally, a hidden layer is sufficient for most analyses, but in highly complex systems, two hidden layers could be used. Each neuron in the hidden layer is connected to all the neurons in its preceding and succeeding layers [48]. The amount of each neuron in the hidden layer and the output layer is determined based on the amount of each neuron in the previous layer, weights, and bias. To do so, the amount of each neuron in the previous layer is multiplied by its weight and then the sum of the weighted values of neurons in the previous layer is obtained and combined with the bias. The obtained value is passed through an activation function and transferred to the next layer [48]. Various activation functions are used in MLP, including Tansig, Linear, Sigmoid and Tanh.

Optimization algorithms used for model training play a key role in MLP performance. In other words, training optimization in a neural network is equivalent to minimizing a general error function, which is a multivariate function and depends on the weights of the network. In this study, LM and SCG algorithms have been used to optimize MLP.

The LM algorithm introduced by Kenneth Levenberg and Donald Marquardt is a simple and stable convergence algorithm, which represents the most prevalent way of optimizing weights and biases in MLP networks [49]. This algorithm is a combination of the steepest descent method and the Gauss-Newton algorithm, which is designed to alleviate computations by excluding the Hessian matrix [50]. Interested readers can refer to [51] for further details regarding the application of the LM algorithm.

Another set of training algorithms for MLP neural networks is the Conjugate Gradient (CG) algorithms, for which a variety of algorithms have been presented so far. In conventional CG algorithms, the step size is estimated using the line search technique, which escalates the computational complexity. The SCG algorithm used in this paper is a CG algorithm that eliminates the line search technique and utilizes a step size scaling mechanism, thus accelerating the network learning process [52,53].

The MLP neural network used in this article has 40 neurons in 4 hidden layers for LM and SCG algorithm. Tansig, Sigmoid, Tansig, and Tansig activation functions, respectively, were used in hidden layers.

3.4.2. Radial Basis Function (RBF) Neural Network

RBF is one of the most popular neural networks introduced by Broomhead and Lowe in 1988. Employed for both classification and regression purposes, this neural network is inspired by approximation function theory. The RBF generally has a three-layer feed-forward architecture in which an input layer connects to the output layer via a hidden layer [54,55]. Figure 4 illustrates the structure of the RBF neural network adopted in this paper.

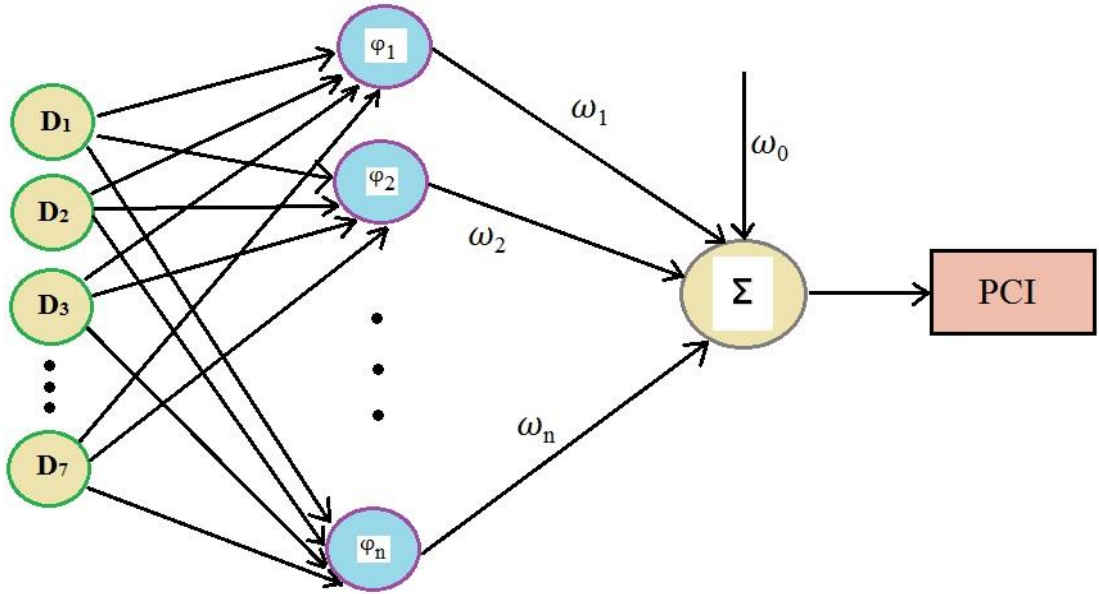

**Figure 4.** Structure of RBF neural network used in this paper

The input layer contains seven nodes (input variables including $D_1$, $D_2$, $D_3$, $D_4$, $D_5$, $D_6$, and $D_7$). The main member of the RBF network is the hidden layer that transfers information from the input layer to the hidden space. Each point in the hidden layer is the center of a specific space with a known radius [56]. For inputs of research ($D_i$), RBF network calculates PCI as Eq.6 [57]:

$$\text{PCI} = \sum_{j=1}^{N} w_j \varphi_j(\|D_i - c_j\|) \quad (6)$$

where, $w_j$: Connection weight, $\varphi_j$: Radial basis function, and $\|D_i - c_j\|$: Euclidian distance between input data and radial function center.

In this paper, GA and ICA algorithms are used to optimize the RBF neural network. GA is a meta-heuristic algorithm inspired by natural selection processes and used for search and optimization problems. In this algorithm, a set of possible solutions, phenotype, is developed for an optimization problem to find better solutions. Each person has a set of chromosomes and genotypes that could be modified or stimulated. In this algorithm, a population of individuals generated in a random process begins to evolve. The fitness (target function value) of each individual in the population is determined and the fittest individuals are selected to produce the next generation. The new generation will be used in the next iteration of the algorithm. This process is sustained until the maximum number of iterations (or the highest number of generations) or the desired accuracy in the optimization problem is achieved [58,59]. Figure 5 shows the GA algorithm.

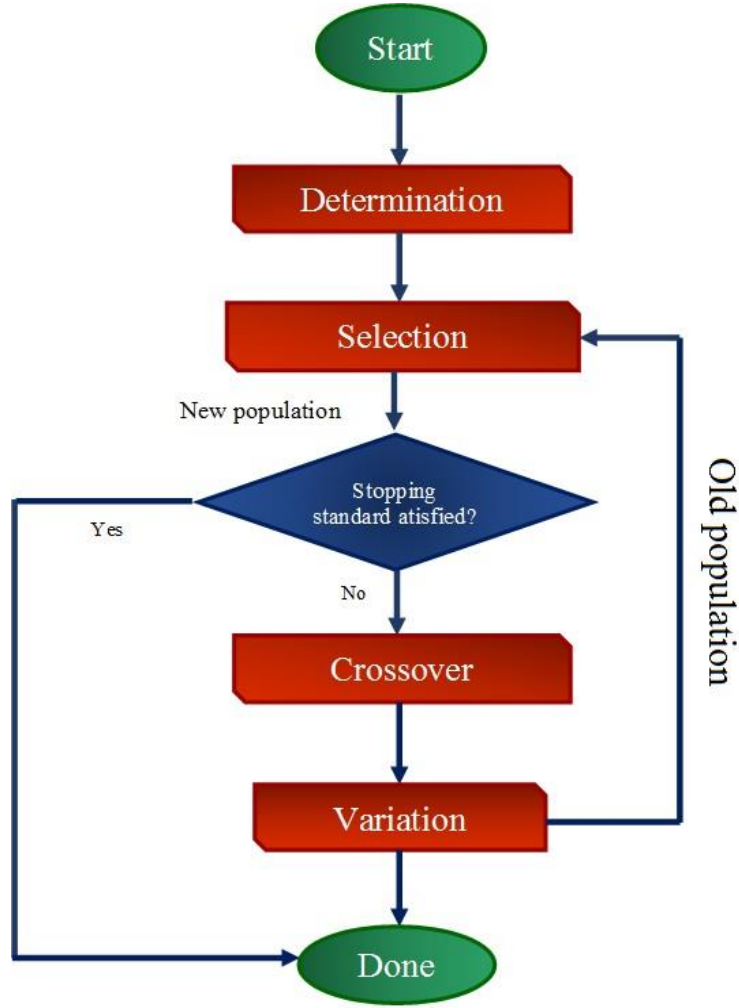

**Figure 5.** A schematic of GA method in this study

ICA is an algorithm inspired by colonial rivalry, representing an evolutionary algorithm for optimization problems. This algorithm was first proposed by Atashpaz-Gargari and Lucas [60]. Like other evolutionary algorithms, ICA begins with an initial population (countries of the world). These countries are split into two categories including imperialist states and colonies. All colonies are divided among the imperialists incommensurate with their power and dominance. Each empire consists of an imperialist and a few colonies. The power of each empire corresponds to the fitness value at the GA algorithm and embraces the power of the colonial state and its colonies [60]. Over time, the colonies begin to launch a movement against the imperialists, and some powerful colonies may be able to seize the power of the empire. In the next stage, a rivalry breaks out between the imperialists with the strong empires gradually growing in strength and the feeble empires collapsing. The movement of colonies against the imperialists, the rivalry of imperialists, and the dissolution mechanism continue until all countries merge into one state with only a single empire while other countries serve as its colonies. Under these conditions, since all colonies are in a relatively identical state and they all have the same position and value, the algorithm ends [60,61]. Figure 6 shows the flowchart of the ICA algorithm.

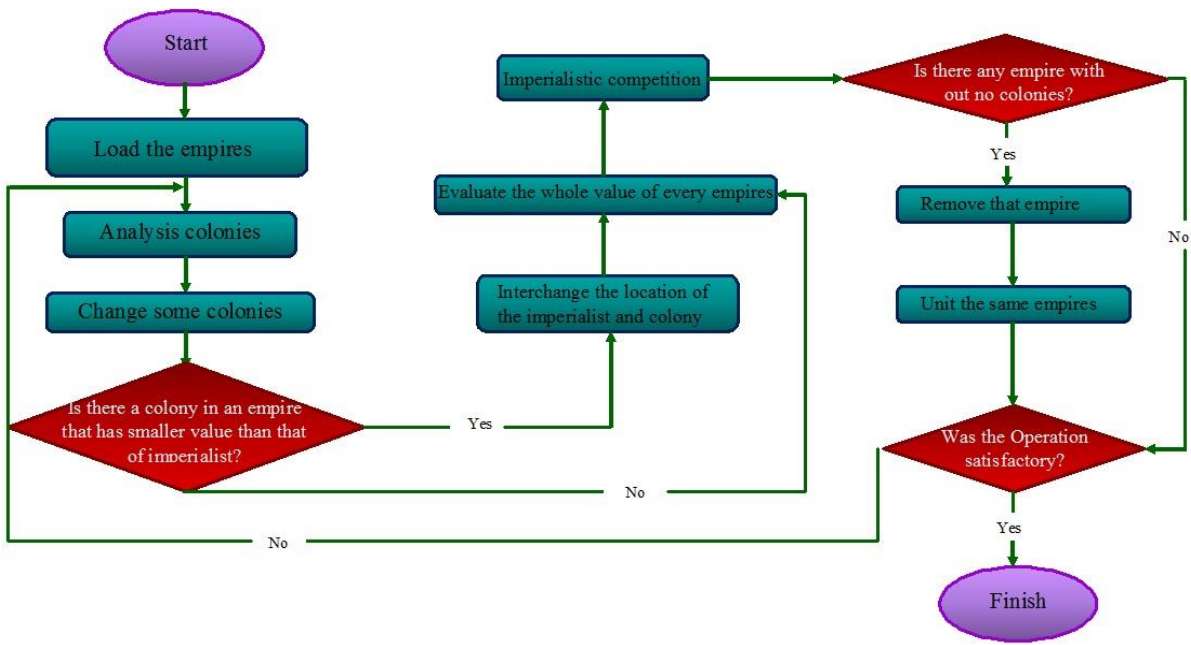

**Figure 6.** Flowchart of ICA

For GA and ICA algorithms used in this study, the number of neurons and the distribution coefficient were 55 and 0.37, respectively.

3.4.3. Committee Machine Intelligent System (CMIS)

The standard procedure in intelligence analysis is to consider several models for analysis and then select the best model based on the results. In this process, efforts made for the abandoned models are virtually in vain. This drawback could be fixed by a committee machine. In a committee machine, the results of different models are combined to reach a more accurate answer. The important thing in a committee machine is how to integrate models. In simple arithmetic averaging, all solutions have the same contribution, but in weighted averaging, the solutions are weighted based on their accuracy and then incorporated into the final solution [62-64]. In this paper, the CMIS model was presented by using MLP-LM, MLP-SCG, RBF-GA, and RBF-ICA neural networks. Weighted coefficients were optimized by Solver. Table 5 shows the final weighted coefficients in CMIS.

**Table 5.** Coefficients of CMIS

| No. of coefficients | Coefficients |
|---|---|
| $C_1$ | 0 |
| $C_2$ | 0.657295 |
| $C_3$ | 0.227583 |
| $C_4$ | 0.069749 |
| $C_5$ | 0.04656 |

*3.5. Performance Criteria*

In any scientific study, after analyzing data and calculating the analysis output, the results need to be reviewed and verified. Four statistical criteria have been used to validate the results in this study including APRE, AAPRE, RMSE, and SD. These criteria are calculated according to the following Eqs. 7 to 10 [65]:

$$\text{APRE} = \frac{100}{N}\sum_{i=1}^{N}\frac{\text{PCI}_{\text{observed},i} - \text{PCI}_{\text{predicted},i}}{\text{PCI}_{\text{observed},i}} \tag{7}$$

$$\text{AAPRE} = \frac{100}{N}\sum_{i=1}^{N}\frac{|\text{PCI}_{\text{observed},i} - \text{PCI}_{\text{predicted},i}|}{\text{PCI}_{\text{observed},i}} \tag{8}$$

$$\text{RMSE} = \sqrt{\frac{1}{N}\sum_{i=1}^{N}(\text{PCI}_{\text{observed},i} - \text{PCI}_{\text{predicted},i})^2} \tag{9}$$

$$\text{SD} = \sqrt{\frac{1}{N-1}\sum_{i=1}^{N}\left[\frac{\text{PCI}_{\text{observed},i} - \text{PCI}_{\text{predicted},i}}{\text{PCI}_{\text{observed},i}}\right]^2} \tag{10}$$

All the above four statistical criteria represent some kind of computational errors, with smaller values close to zero indicating higher accuracy of the modeling results. By examining Eqs. 7 to 10, it becomes clear that APRE can be negative and the other three criteria are always positive.

## 4. Results and Discussion

This section presents and discusses the achieved results. As mentioned in the previous section, five methods MLP-LM, MLP-SCG, RBF-GA, RBF-ICA, and CMIS were applied to the PCI prediction. The modeling input in this study is the pavement surface deflection, which is collected by FWD. Figure 7 shows the relative impact of recorded deflections on PCI. In this figure, $D_1$ to $D_7$ represents the deflections in geophones 1 to 7, respectively. As shown in Figure 7, geophones 1 to 3 is inversely related while other geophones are directly related to PCI. The deflections in geophone 7, which is the furthest from the loading spot, wield the highest impact on PCI.

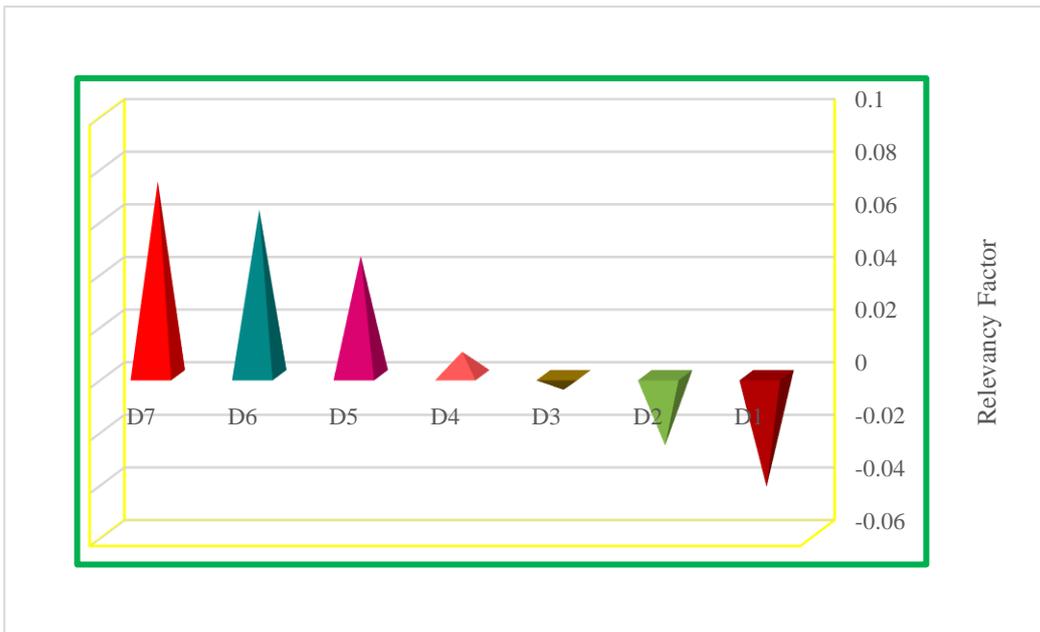

**Figure 7.** The relative effect of input parameters on PCI

Table 6 lists the statistical parameters of APRE, AAPRE, RMSE, and SD for all the models developed in this paper. Considering AARPE and APRE values for the CMIS model (11.67% and 2.33%, respectively) and lower RMSE and SD errors in the CMIS method compared to other models, this model yields the highest accuracy for predicting PCI among the developed models.

Table 6. Performance criteria of the all developed models for prediction of PCI

| Model | Data | APRE (%) | AAPRE (%) | RMSE | SD |
|---|---|---|---|---|---|
| CMIS | Train | 3.5636 | 11.6098 | 12.0543 | 0.020082 |
| | Test | -2.632 | 11.9464 | 11.807884 | 0.025687 |
| | Total | 2.3303 | 11.6768 | 12.005653 | 0.021081 |
| MLP-LM | Train | -1.2093 | 14.9275 | 14.541231 | 0.078995 |
| | Test | 4.7599 | 12.7179 | 14.330003 | 0.027158 |
| | Total | -0.02115 | 14.4877 | 14.499431 | 0.068174 |
| MLP-SCG | Train | -0.1949 | 15.5046 | 14.167214 | 0.070149 |
| | Test | 6.6833 | 13.5662 | 15.145498 | 0.031747 |
| | Total | 1.1742 | 15.1187 | 14.367255 | 0.062318 |
| RBF-GA | Train | -0.4446 | 11.8509 | 13.003455 | 0.032987 |
| | Test | 14.2997 | 19.406 | 58.919147 | 0.360989 |
| | Total | 2.4902 | 13.3547 | 28.747779 | 0.096868 |
| RBF-ICA | Train | -0.5063 | 12.6392 | 17.077364 | 0.057202 |
| | Test | 16.7628 | 25.886 | 45.214386 | 0.45286 |
| | Total | 2.9311 | 15.276 | 25.308415 | 0.134177 |

According to Figure 8, one can visually analyze the quality of all models developed to predict PCI of asphalt pavement. In this figure, a graph is presented for all five models proposed in this research. In each graph, the horizontal axis represents $PCI_{observed}$ values and the vertical axis represents $PCI_{predicted}$ values. In Figure 8, the higher is the data concentration around the Y=X line, the higher is the accuracy of the model in predicting PCI. As can be seen in this figure, the concentration of points around Y=X line for the CMIS model is higher than other models, so this model has greater precision in predicting PCI.

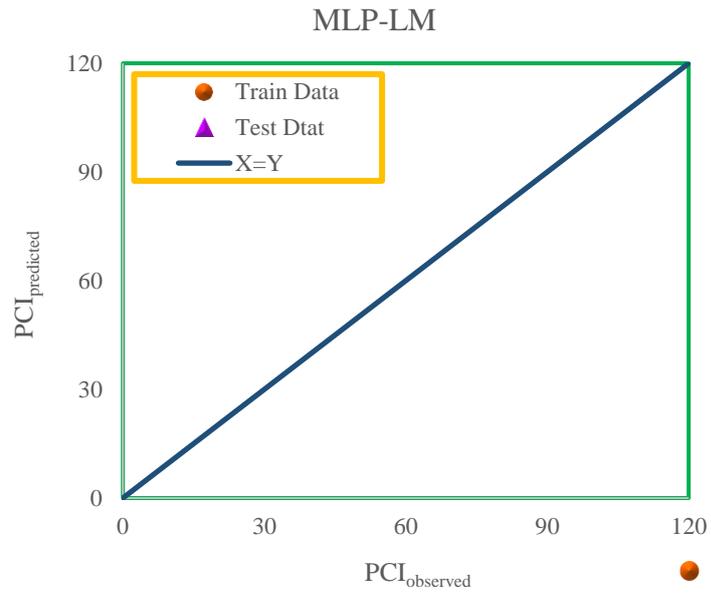

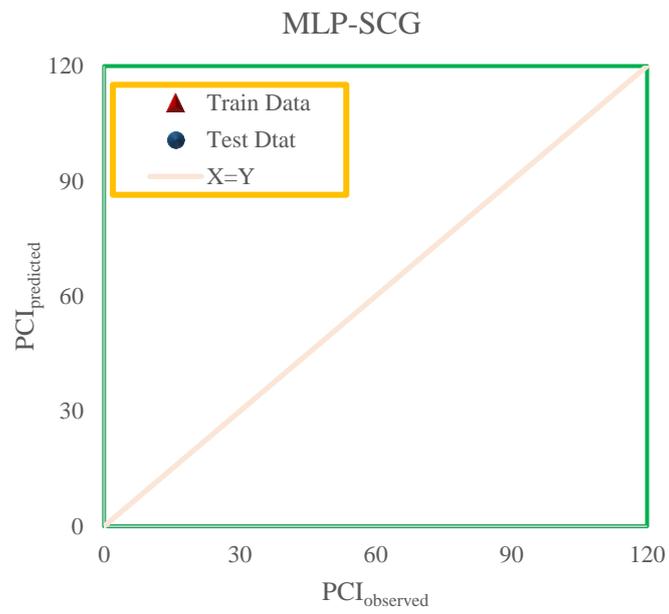

**Figure 8.** Cross-plot for developed models to prediction of PCI

### RBF-ICA

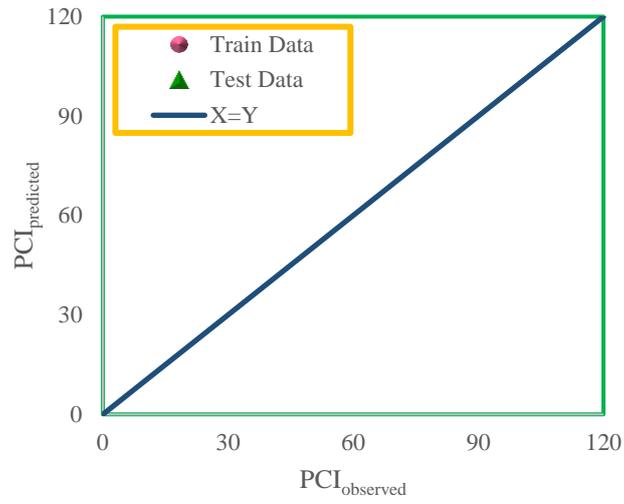

### RBF-GA

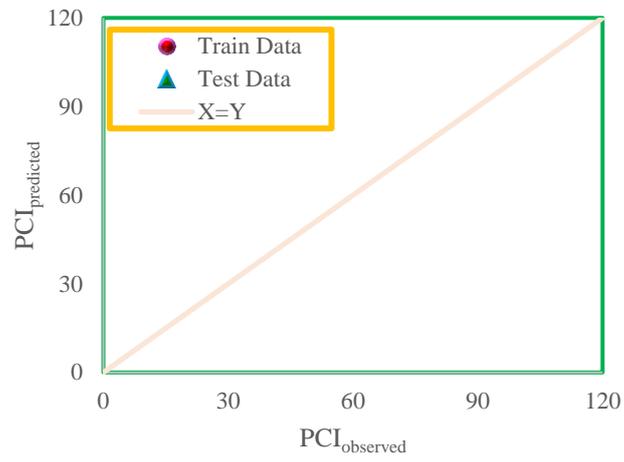

### CMIS

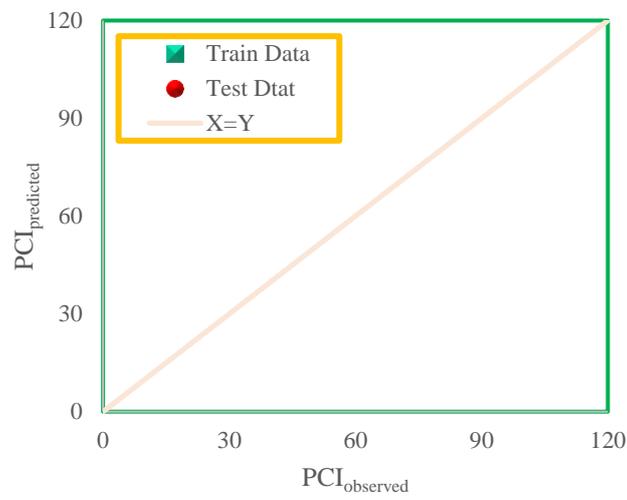

**Figure 8.** Continued

Figure 9 reveals the relative error distribution curves. In general, and for each model, the closer the data points are to the horizontal line of zero error, the greater the accuracy of the model is. According to Figure 9, the highest relative error between PCI$_{observed}$ and PCI$_{predicted}$ in the CMIS model is less than 33%, which is superior to other models. Thus, Figure 9 also confirms the greater accuracy of the CMIS model compared to other models.

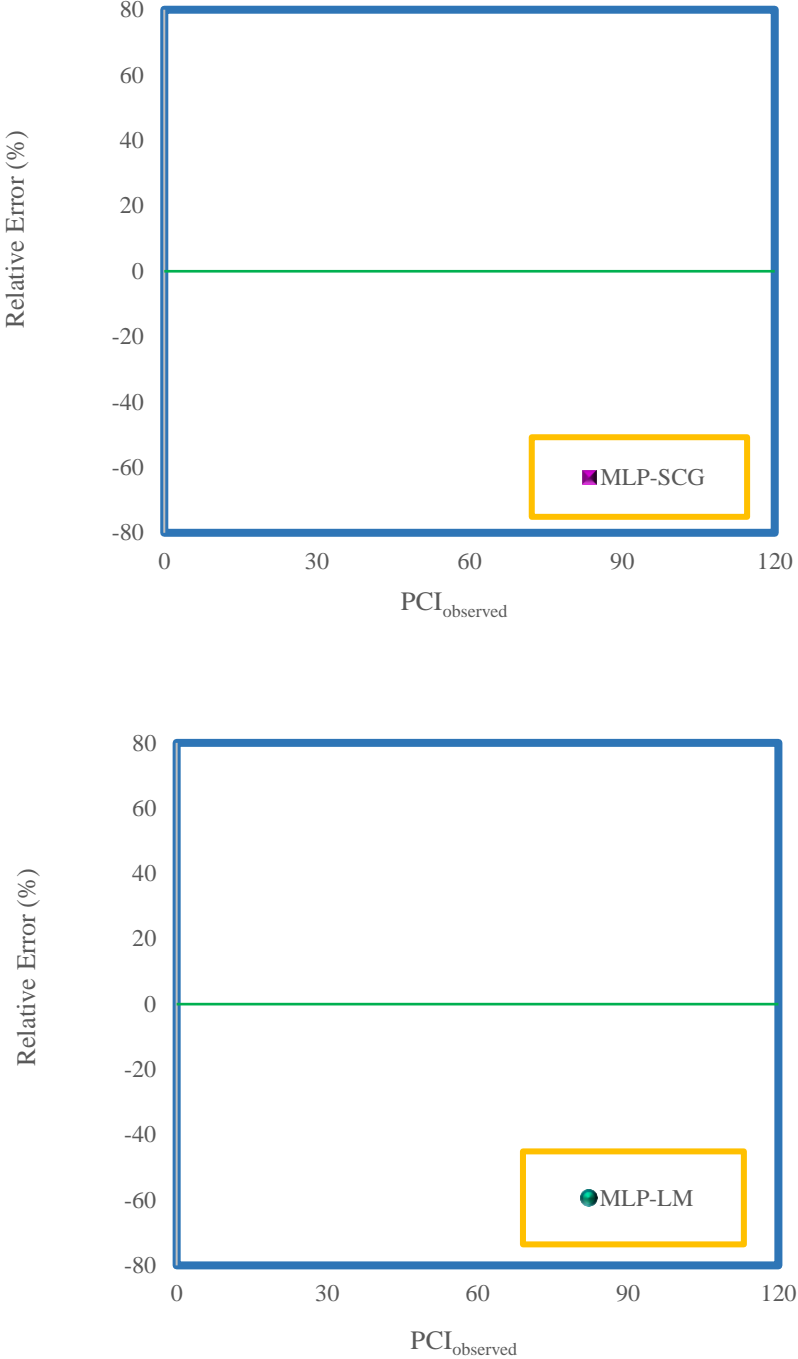

**Figure 9.** Relative error between the observed and predicted PCI versus observed PCI

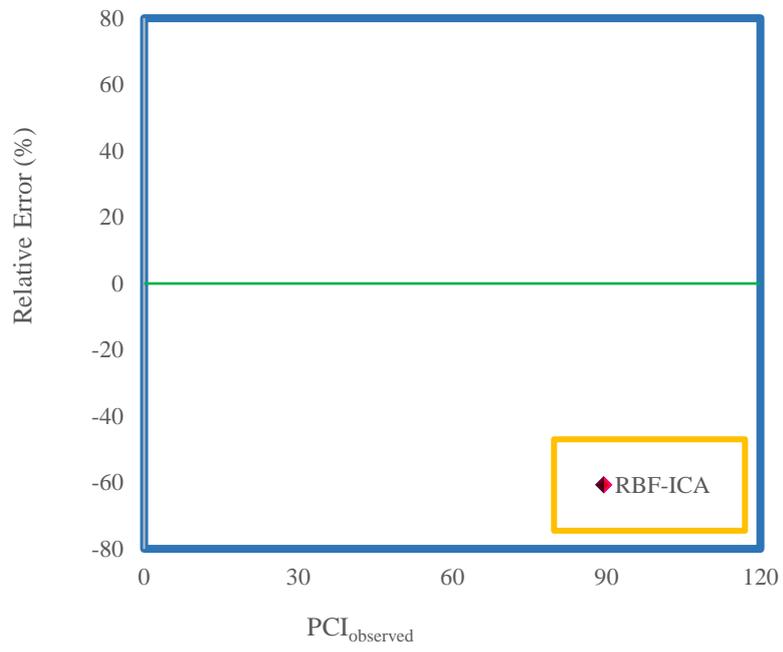

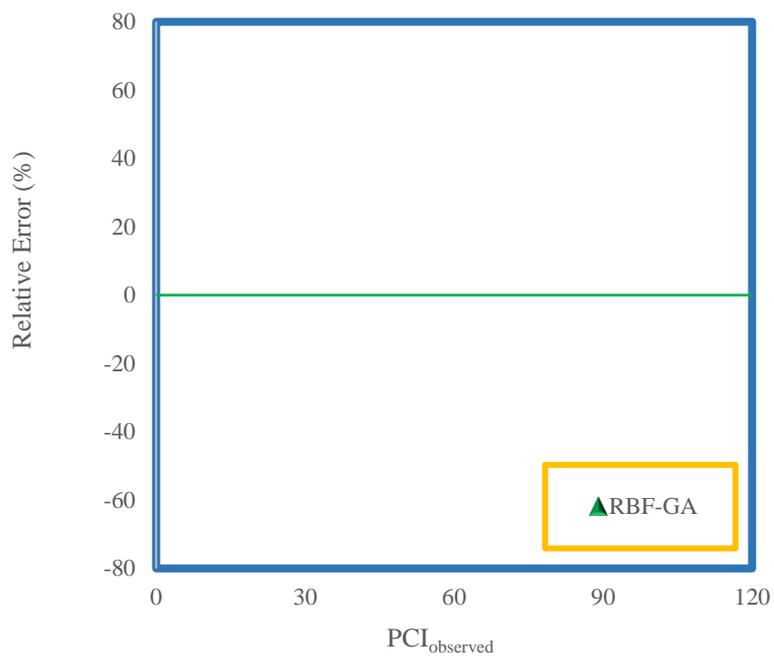

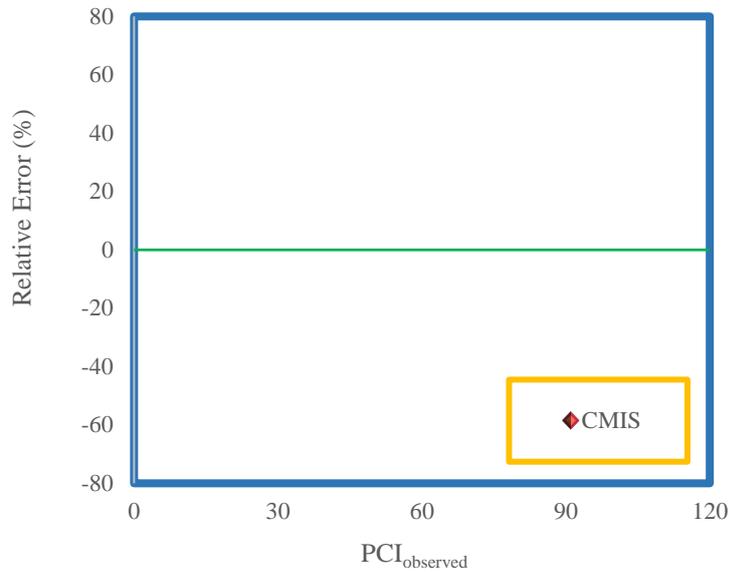

**Figure 9.** Continued

The cumulative frequency curve of AAPRE for models developed in this paper is presented in Figure 10. The analysis of curves in this figure suggests that the quality of results in models based on the RBF neural network is higher than models based on MLP neural network, especially RBF-GA model, which has a lower error in PCI prediction. The curve of the CMIS model, which is the median of the results achieved from four MLP-LM, MLP-SCG, RBF-GA and RBG-ICA methods, lies in the middle of these four methods. However, the endpoint of the CMIS curve is in a better position than all four methods, which corroborates the higher quality of the CMIS model in PCI prediction.

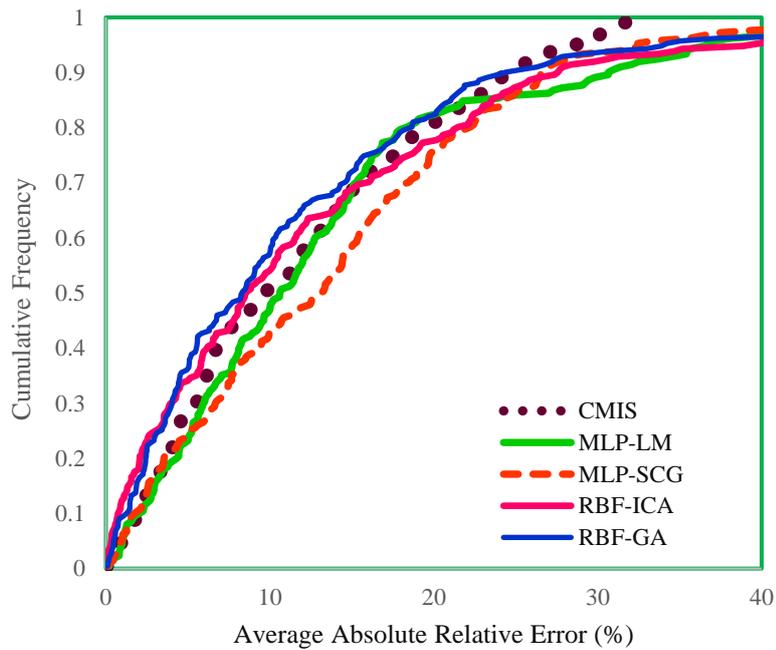

**Figure 10.** Cumulative frequency curve of average absolute relative error for developed models in this study to predict PCI

## 5. Conclusion

In this paper, the authors attempted to present a method for estimation the PCI based on pavement surface deflections in flexible pavements. To implement the proposed theory in this paper, data set including PCI and pavement surface deflections were collected based on FWD testing of 236 pavement segments taken from Tehran-Qom freeway in Iran. The data set was analyzed by two MLP and RBF neural networks. LM and SCG algorithms for optimization of MLP neural network and ICA and GA algorithms for optimization of RBF neural network were used. To improve the results of four neural networks adopted in this study, the CMIS method was employed. The results of this paper were verified by four statistical criteria including APRE, AAPRE, RMSE, and SD. For CMIS method, the values of these criteria were 2.3303, 11.6768, 12.0056 and 0.0210, respectively.

The proposed method in this paper helps pavement engineers to use the non-destructive test (FWD) results for determining PCI rather than a visual survey by the inspector. Therefore, the challenges of the traditional procedure for PCI calculation (safety and potential human error) are eliminated. On the other hand, since FWD is generally used in pavement network maintenance programs, the method proposed provides overlapping in pavement maintenance activities and thus saving time and expense.

For future research, the authors suggest that modeling becomes more complete. For instance, the results of other non-destructive tests such as GPR are also used. On the other hand, after approving in the study phase, new deflectometers (such as RWD and TSD) can be used for recording the pavement surface deflections. Because of having the speed of traffic, these equipment have less interference in traffic flow.

**Acronyms**

| Abbreviation | Description |
| --- | --- |
| PCI | Pavement Condition Index |
| FWD | Falling Weight Deflectometer |
| MLP | Multi-layer perceptron |
| RBF | Radial Basis Function |
| MLP-LM | Multi-layer perceptron Optimized by Levenberg-Marquardt Algorithm |
| MLP-SCG | Multi-layer perceptron Optimized by Scaled Conjugate Gradient Algorithm |
| RBF-ICA | Radial Basis Function Optimized by Imperialist Competitive Algorithm |
| RBF-GA | Radial Basis Function Optimized by Genetic Algorithm |
| CMIS | Committee Machine Intelligent Systems |
| APRE | Average Percent Relative Error |
| AAPRE | Average Absolute Percent Relative Error |
| RMSE | Root Mean Square Error |
| SD | Standard Error |
| MR&R | Maintenance, Rehabilitation, and Reconstruction |
| PMS | Pavement Management System |
| ML | Machine Learning |
| DV | Deduct Value |
| TDV | Total Deduct Value |
| CDV | Corrected Deduct Value |
| CG | Conjugate Gradient |

**Author Contributions:** Conceptualization, Shahab Shamshirband, Amir Mosavi and Peter Csiba; Data curation, Nader Karballaeezadeh and Farah Zaremotekhases; Formal analysis, Nader Karballaeezadeh, Farah Zaremotekhases, Narjes Nabipour and Várkonyi-Kóczy Annamária; Funding acquisition, Amir Mosavi and Várkonyi-Kóczy Annamária; Investigation, Nader Karballaeezadeh, Farah Zaremotekhases and Peter Csiba; Methodology, Shahab Shamshirband and Amir Mosavi; Project administration, Várkonyi-Kóczy Annamária; Resources, Shahab Shamshirband, Narjes Nabipour, Peter Csiba and Várkonyi-Kóczy Annamária; Software, Nader Karballaeezadeh, Farah Zaremotekhases, Shahab Shamshirband and Narjes Nabipour; Supervision, Peter Csiba and Várkonyi-Kóczy Annamária; Validation, Amir Mosavi, Narjes Nabipour and Peter Csiba; Visualization, Narjes Nabipour; Writing – original draft, Farah Zaremotekhases and Amir Mosavi.

**Acknowledgments:** We acknowledge the support of this paper within the project of "Support of research and development activities of the J. Selye University in the field of Digital Slovakia and creative industry" of the Research & Innovation Operational Programme (ITMS code: NFP313010T504) co-funded by the European Regional Development Fund.

**Conflicts of Interest:** The authors declare no conflict of interest.